\documentclass[journal,twoside,web]{ieeecolor}
\usepackage{jsen}
\usepackage{cite}
\usepackage{amsmath,amssymb,amsfonts}
\usepackage{algorithmic}
\usepackage{graphicx, caption, subcaption}
\usepackage{textcomp}
\usepackage{wrapfig}
\def\BibTeX{{\rm B\kern-.05em{\sc i\kern-.025em b}\kern-.08em
		T\kern-.1667em\lower.7ex\hbox{E}\kern-.125emX}}
\definecolor{abstractbg}{rgb}{0.89804,0.94510,0.83137}
\begin{document}
	\title{A Low-cost Through-metal Communication System for Sensors in Metallic Pipes}
	\author{Hongzhi Guo, \IEEEmembership{Senior Member, IEEE}, Marlin Prince, Javionn Ramsey, Jarvis Turner, Marcus Allen, \\Chevel Samuels, and Jordan Atta Nuako
		\thanks{This work was supported in part by the U.S. National Science Foundation under Grant HRD1953460 and CNS1947748, and The Thomas F. and Kate Miller Jeffress Memorial Trust, Bank of America, Trustee. }
		\thanks{Hongzhi Guo is with the School of Computing, University of Nebraska-Lincoln, NE, 68588, USA. (e-mail: hguo10@unl.edu). M. Prince, J. Ramsey, J. Turner, M. Allen, C. Samuels, and J. A. Nuako are with the Engineering Department, Norfolk State University, VA, 23504, USA.}
	}
	
	\IEEEtitleabstractindextext{%
		\fcolorbox{abstractbg}{abstractbg}{%
			\begin{minipage}{\textwidth}%
				\begin{wrapfigure}[12]{r}{3in}%
					\includegraphics[width=3in]{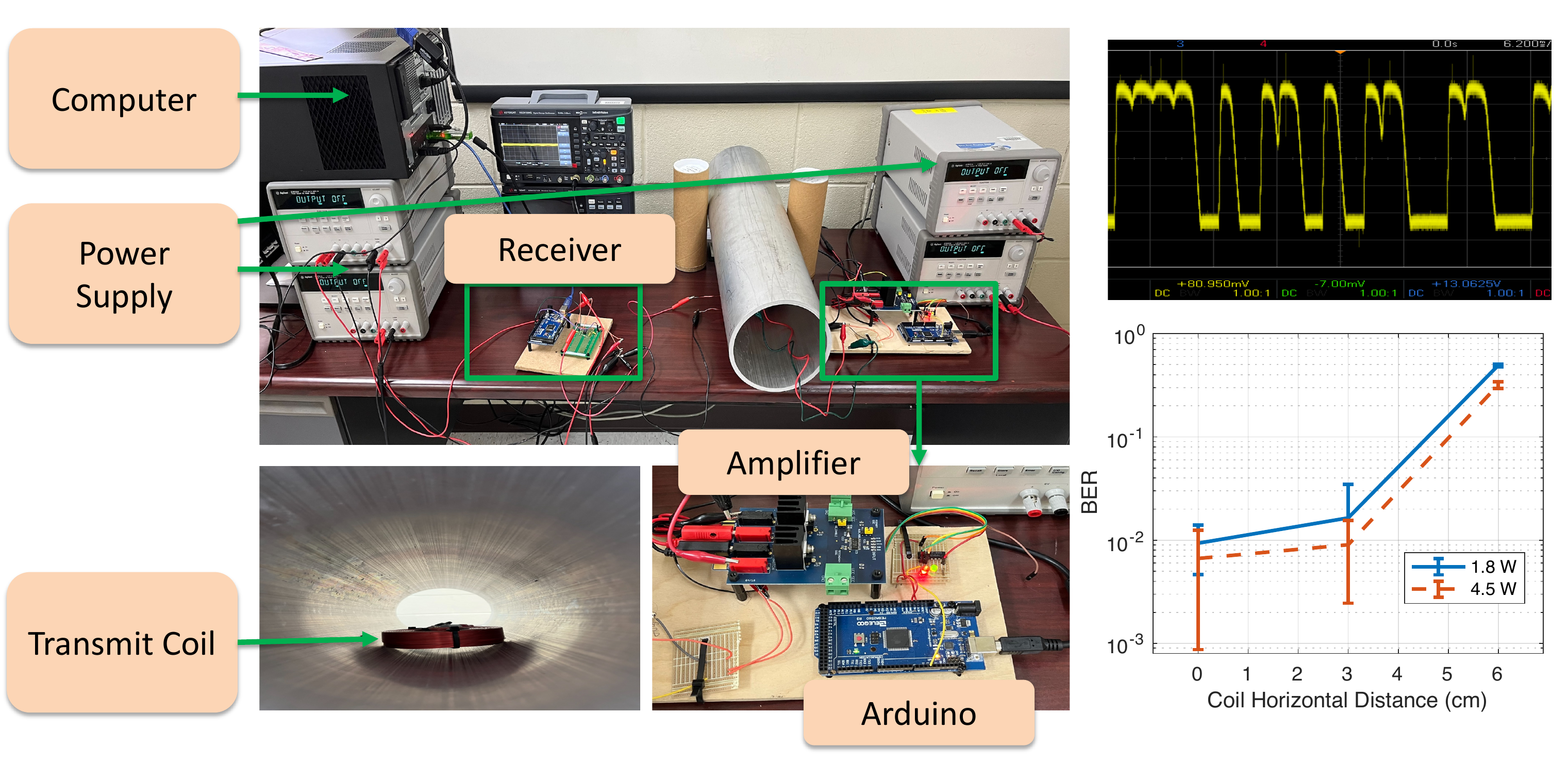}%
				\end{wrapfigure}%
				\begin{abstract}
					Metallic pipes and other containers are widely used to store and transport toxic gases and liquids. Various sensors have been designed to monitor the environment inside metallic pipes and containers, such as pressure, liquid-level, and chemical sensors. Moreover, sensors are also used to inspect and detect pipe leakages. However, sensors are usually placed outside of metallic pipes and containers and use ultrasound to monitor the internal unseen environment. This is mainly due to the fact that internal sensors cannot communicate with external data sinks without cables, but using cables can dramatically affect the metal-sealed structure. Wireless communication is desirable to communicate with internal sensors, but it experiences high attenuation losses since metal can block wireless signals due to its high conductivity. This paper develops a low-cost through-metal communication system prototype using off-the-shelf electronic components. The system is fully reconfigurable, and arbitrary modulation and coding schemes can be implemented. We design the transmit module which includes a signal processing microcontroller, an amplifier, and a transmit coil, and the receive module which includes a receive coil, an amplifier, and a microcontroller with demodulation algorithms and bit-error-rate (BER) calculations. The performance of the prototype is evaluated using various symbol rates, distances, and transmission power. The results show that the communication system can achieve a 500 bps data rate with 0.01 BER and 3.4 cm communication range when penetrating an Aluminum pipe with 7 mm thickness.   
				\end{abstract}
				
	\end{minipage}}}
	
	\maketitle
	
	\section{Introduction}
	\label{sec:introduction}
	\IEEEPARstart{M}{etallic} pipes and other containers are used to store or transport toxic and corrosive liquids and gases. They are tightly sealed to avoid leakages which can generate significant economic losses and contaminations. Various technologies have been proposed to monitor and predict leakages of metallic containers and pipes, such as wireless sensor networks \cite{sun2011mise}, in-pipe robots \cite{qi2010wireless}, and, more recently, digital twins \cite{wanasinghe2020digital}. All these technologies rely on the employment of sensors to persistently monitor metallic containers and pipes. 
	
	Sensors have been employed in metal-constrained environments to gather critical information. Giant Magnetoresistance (GMR) sensors can be used to find flaws in metallic pipes \cite{pelkner2018benefits}. Due to their exceptional sensitivity to magnetic flux leakage, even minor defects can be identified. Additionally, changes in the internal condition of a metallic pipeline, such as cracks, have an impact on the flow and pressure of the contents it transports. Sensors monitoring flow and pressure can aid in the detection of these variations in the substance's flow and pressure patterns.
	Camera sensors can also be deployed in metallic pipelines to detect defects. In \cite{el2011vison}, an automated 3D pipe reconstruction system using a single off-the-shelf camera as the only sensor in the system is presented. As part of the experiment, defects in metallic pipes are detected by identifying the presence of a cluster of outliers from the images captured by sensors which are fed to a cylinder fitting algorithm. Ultrasonic sensors are tools that can measure the surface thickness of a pipeline. Based on the measurements, mechanical defects on the internal surface of a metallic pipeline can be detected \cite{alnaimi2015design}. 
	
	For metallic containers, such as the metal structure of aircraft, eddy current sensors can measure a crack with an accuracy of 1 mm and an average error of 4.6\% compared with fracture analysis \cite{jiao2016monitoring}. Metallic containers can be fitted with liquid-level sensors to track the liquid level continuously. A liquid-level sensor based on surface plasmon resonance (SPR) can detect each interface, such as the fuel-air and fuel-water interfaces, and can measure the level of each liquid in real-time in the oil industries where water, gasoline, and air are typically in the same tank \cite{pozo2016continuous}. Furthermore, temperature sensors are also widely used since it is crucial to pay close attention to how the temperature is distributed when monitoring the health of oil and gas storage containers. An application is presented in \cite{fan2015large} and a wireless temperature monitoring system for Liquified Petroleum Gas (LPG) is developed. Nine sensors are installed on three spherical LPG storage tanks to monitor the internal temperature. The temperatures in the containers are efficiently provided by sensors at all sensing points. In \cite{kar2021passive}, ultrasonic temperature sensors are used to measure temperature in a sealed metallic container.

	Most of the existing sensors for monitoring metallic containers and pipes are placed outside of the structure. This is mainly due to the challenge of wireless communication through the metal wall \cite{shan2020developing,guo2019reliable}, where the high electric conductivity of metal blocks wireless signals. In \cite{yan2020deep}, ultrasonic sensors are placed on the external surface of gas pipes to detect metal cracking. In \cite{yang2021bilstm}, a theoretical model is developed to estimate and localize pipeline leakages based on external sensors. However, external sensing has limited capability which is not comparable to internal sensing which can provide mechanical, chemical, temperature, and many other information of interest. Wired communications for sensors inside metallic containers or pipes require holes to route wires. This reduces the tightness and mechanical strength of the metallic structure. Also, static sensors that are installed at fixed locations cannot actively inspect leakages. In-pipe robots can actively inspect a metallic pipeline. In \cite{waleed2018pipe}, a mobile robotic sensor is developed to inspect acrylic pipes with leakage detection function using deep learning models. However, most existing in-pipe robots are tethered using cables to provide power and communication support, which limits the range and flexibility of deployment. Although challenging, wireless communication technologies are desirable for sensors and robots inside metallic containers and pipes.
	
	There are three wireless communication techniques for through-metal applications, namely, acoustic communication \cite{ashdown2018high}, capacitive coupling \cite{erel2021comprehensive}, and magnetic inductive coupling and magnetic resonance \cite{guo2019reliable}. Acoustic communication is contact-based which can achieve high efficiency in wireless power transfer and high data rates in wireless communication. Acoustic communication cannot be efficiently implemented for mobile robotic sensors due to the contact with metal walls. The capacitive coupling using two plates has been used for wireless power transfer through metal walls, but its efficiency is not clear and it is barely used for data transmission. Magnetic resonance uses resonant coils that can achieve high efficiencies in wireless power transfer through metal walls. However, due to the resonance of coils, the bandwidth of magnetic resonance is narrow, which is not efficient in high-data-rate wireless communication \cite{yang2015through}. 
	
	\begin{table*}
		\begin{center}
			\caption{Existing Through-metal Wireless Power Transfer and Wireless Communication Prototypes.}
			\label{tab:comparison}
			\begin{tabular}{||c| c |c |c|c|c||} 
				\hline
				Ref. & Objective & Metal Thickness& Frequency & Efficiency & Data Rates/Bandwidth\\ [0.5ex] 
				\hline\hline
				\cite{zangl2010wireless}& WPT& Stainless Steel, 1.6 mm & 50 Hz &  &  \\ 
				\hline
				\cite{zangl2010wireless} &WC &Stainless Steel, 1.6 mm & 49 kHz &  & 2 kHz \\
				\hline
				\cite{romero2021miniature} & WPT & Aluminum, 1 mm & 2 kHz& 1.46 \% &  \\
				\hline
				\cite{yamakawa2014wireless} & WPT & Stainless Steel, 1 mm & 50 Hz & 43\%& \\
				\hline
				\cite{yamakawa2014wireless} & WPT & Metal Plate, 5 mm & 50 Hz & 10\%&\\  
				\hline
				\cite{qi2010wireless} & Localization & Metallic pipe, 12 mm & 23.4 Hz & &\\  
				\hline
				\cite{van2017development} & WPT & Aluminum, 3 mm & 980 Hz & 3.7\%&\\
				\hline
				This paper & WC & Aluminum, 7 mm & 500 Hz & & 500 bps\\
				\hline
			\end{tabular}\\
			\vspace{4pt}
			\footnotesize{Note: WPT stands for Wireless Power Transfer and WC stands for Wireless Communication. }
		\end{center}
	\end{table*}
	
	We use magnetic induction for through-metal wireless communication in this paper. Although previous work has demonstrated that magnetic induction can be used for both through-metal wireless power transfer and wireless communication, the following research challenges have not been addressed. First, the optimal frequency is determined by the tradeoff between the magnetic coupling and attenuation loss. Existing works use different frequency bands ranging from 23.4 Hz \cite{qi2010wireless} to 49 kHz \cite{zangl2010wireless}, as shown in Table~\ref{tab:comparison}. The optimal frequency should be determined by the metal wall thickness and communication data rates. Existing through-metal wireless power transfer uses a narrow bandwidth which cannot be adopted for wireless communication. A high frequency of 49 kHz was used in \cite{zangl2010wireless} due to the small thickness of the metal wall. Considering the low-frequency bands, the bandwidth must be efficiently utilized to increase the data rate. Second, low-cost design using off-the-shelf components is desirable. Existing works use wireless communication equipment which prevents the accessibility of through-metal wireless system design, test, and implementation. For example, in \cite{qi2010wireless} the amplifier has a gain of 4,800 which cannot be easily obtained using off-the-shelf low-cost electronics. Also, from Table~\ref{tab:comparison}, we notice that there are limited works using magnetic induction/resonance for wireless communication.       
	
	In this paper, we design a magnetic induction communication system for through-metal applications {using off-the-shelf components. {The contribution of this paper is two-fold. 
			\begin{itemize}
				\item First, we analyze the system's optimal frequency and design the coil, modulation schemes, transmission circuits, receiving circuits, and demodulation schemes. Our design choices and underlying analyses are provided.
				\item Second, we assemble the communication system and test its performance in metal-constrained environments, i.e., an Aluminum pipe. The impact of communication range, data rate, and transmission power are analyzed.
		\end{itemize}  }
		The major novelty of this paper is the low-cost design of the reconfigurable communication system. The system design, including code, off-the-shelf components, and circuits, and the collected data in system evaluation are made available in \cite{website}. Although this system is used for through-metal communication, the signals that are transmitted and received can also be used for various through-metal sensings, such as metal thickness estimation and metal wall crack sensing.  }
	
	The rest of this paper is organized as follows. In Section II, we obtain the optimal frequency and study the impact of metal on the coil parameters. In Section III, we introduce the transmitter and the receiver design, including modulation, transmitter amplifier, receiver amplifier, analog signal sampling, and demodulation. After that, we evaluate the design in Section IV. This paper is concluded in Section V.

	\section{Optimal Frequency and Impact of Metal}
	\label{sec:optimal}
	The optimal frequency is determined by the thickness of the metallic pipe. According to Maxwell's equation, the electric field is strongly affected by the metal pipe due to its high conductivity, while static magnetic fields can penetrate metal. However, this is only valid when the operating frequency is nearly zero. As the frequency increases, the electric fields and magnetic fields are coupled together to form electromagnetic waves. As a result, using high frequency, both the electric and magnetic fields can be blocked by the metal wall. Skin depth is a simple parameter that can be used to determine the maximum frequency, which is $\delta = \sqrt{2/\sigma\omega\mu}$, where $\sigma$ is the metal conductivity, $\omega=2\pi f_c$ is the angular frequency, and $f_c$ is the signal carrier frequency, and $\mu$ is the permeability. According to Faraday's law, the voltage induced in the receiver's coil is proportional to the time derivative of the magnetic flux $\phi$. The magnetic flux $\phi=\mu S_r H_r e^{j\omega t}$, where $H_r$ is the magnetic field at the receiver and $S_r$ is the area of the receiver coil. Here, by using approximations, we have $H_r\approx H_t e^{-d/\delta}$, where $H_t$ is the magnetic field outside of the metallic pipe and $d$ is the thickness of the metallic pipe. Note that, the magnetic field in the near field falls off fast even without metal blockages. We do not consider this effect since the metal absorption loss is much more significant. Since $H_r$ is proportional to the received voltage and $H_t$ is proportional to the transmitted voltage, the voltage drop can be estimated using $e^{-d/\delta}$. Given the transmitted coil voltage $V_t$, the desired received voltage $V_r$, and the metallic pipe thickness $d$, the maximum carrier frequency is 
	\begin{align}
		\label{equ:fc}
		f_c \leq \frac{(\log V_t - \log V_r)^2}{\pi d^2\sigma \mu}. 
	\end{align}   
	For example, in our prototype, $V_t=30$ V, $V_r=0.2$ V, and $d=7$ mm. The metal material is Aluminum with considered conductivity of 3.5$\times10^7$ S/m. As a result, $f_c\leq 3.7$ kHz. This means higher frequency cannot efficiently penetrate the metallic pipe. Since using a lower frequency reduces the communication bandwidth and the coupling strength between the transmit coil and the receive coil, we prefer using frequencies around 3.7 kHz. However, due to other constraints, the practical frequency is usually much lower than it, which will be discussed in Section \ref{sec:numerical}. Note that, for other applications using magnetic induction, e.g., through soil \cite{liu2021data}, the frequency can be much higher (several MHz) due to the long skin depth. 
	
\begin{figure}
		\centering
		\begin{subfigure}[b]{0.247\textwidth}
			\centering
			\includegraphics[width=\textwidth]{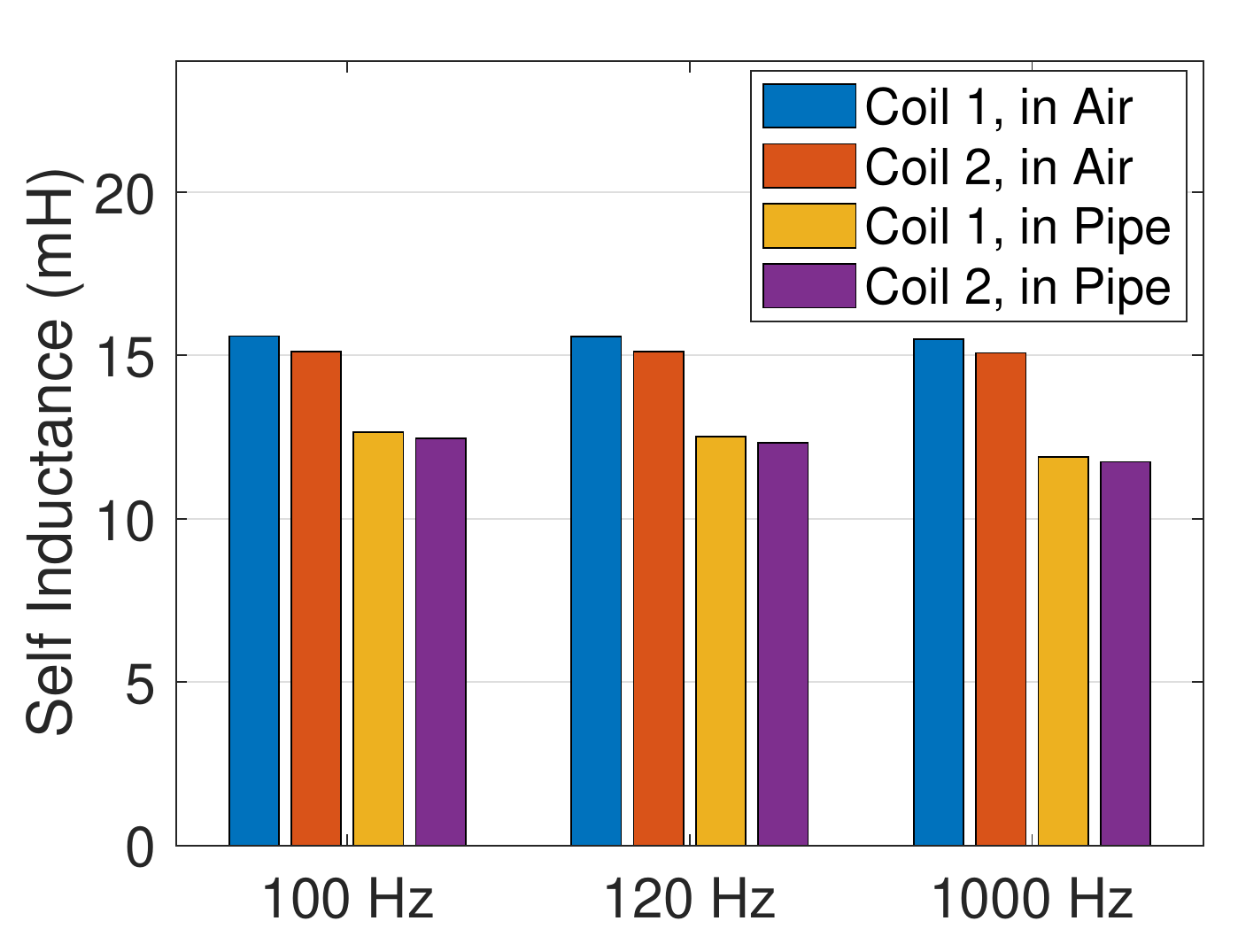}
			\caption{Self-inductance. }
			\label{fig:inductance}
		\end{subfigure}
		\hfill
		\begin{subfigure}[b]{0.233\textwidth}
			\centering
			\includegraphics[width=\textwidth]{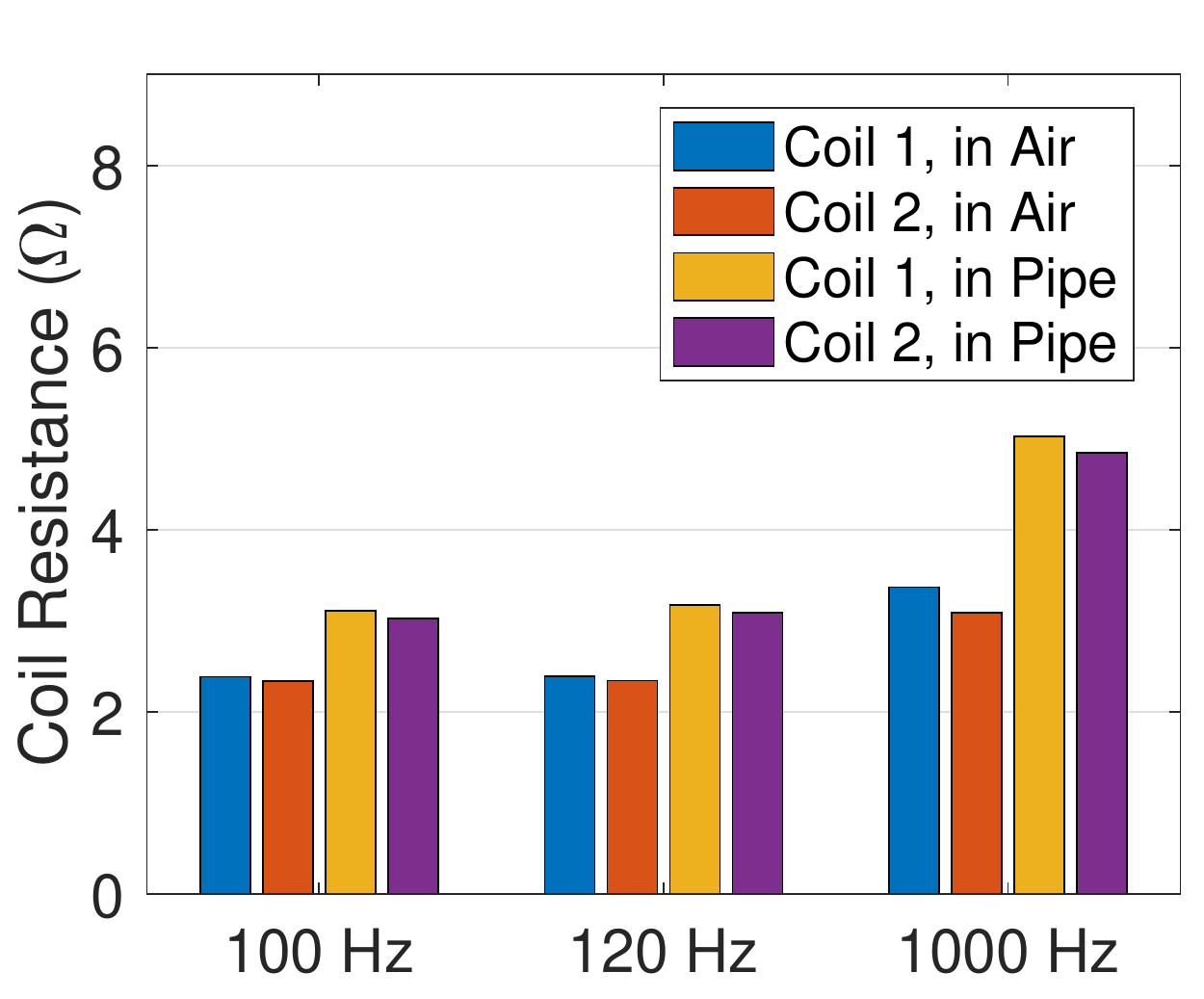}
			\caption{Resistance.}
			\label{fig:resistance}
		\end{subfigure}
		\caption{Coil self-inductance and resistance in the air without metallic pipe and in the metallic pipe. }
		\label{fig:coilparameter}
	\end{figure}

	In Fig.~\ref{fig:coilparameter}, we measured the coil self-inductance and coil resistance in two scenarios, i.e., in the air without metal walls and inside a metallic pipe. The metallic pipe is made of Aluminum with 609.6 mm in length, 7 mm in thickness, and 168 mm in diameter. We use two ERSE 15mH 18 AWG Perfect Layer Inductor Crossover Coils (\$47.49 each). The coil has a diameter of 95 mm. An RSR LCR Meter model 2821 is used to measure the self-inductance and resistance. The coil is placed in the middle of the pipe when measuring its parameters in the pipe. As shown in Fig.~\ref{fig:coilparameter}, the two coil's self-inductance decreases from around 15 mH (in the air) to 12 mH (in the pipe), while the resistance increases, e.g., coil 1's resistance at 1000 Hz increases from 3.369 $\Omega$ in the air to 5.025 $\Omega$ in the pipe. The change is due to the coupling between the metal wall and the coil, which creates reflected impedance. The impedance reduces the coil's self-inductance but increases its resistance \cite{guo2019reliable}.

	Although the impedance change is not significant, it can generate significant impacts on communication performance. To create an LC resonant circuit, for a 15 mH coil, we need a capacitor of 6.75 $\mu$F to achieve resonance at 500 Hz. However, if we place the coil in a metallic pipe, the coil self-inductance is reduced. In Fig.~\ref{fig:resonance}, we show that with such a capacitor and different coils, the absolute values of the overall impedance (in series) demonstrate different resonant frequencies. For example, if we design the coil resonant at 500 Hz in the air, when we place it in the pipe, the resonant frequency is increased (see the response of 12 mH). If the coil still operates at 500 Hz, the efficiency is reduced. Also, we notice that the bandwidth of the resonant circuit is extremely narrow. This means that signals with wider bandwidth can be distorted by the coil, and the information in the signal's amplitude and phase is lost. The receiver cannot demodulate received signals. Due to the narrow bandwidth of the coil, we remove the capacitor and only use magnetic induction for through-metal communication. This allows wideband signals and the data rate can be 500 bps or even higher. Also, since there is no resonant frequency without the capacitor,  the change of the self-inductance due to the metallic pipe does not distort transmitted and received signals. As we can see in Fig.~\ref{fig:resonance}, for 12 mH and 15 mH, the impedance change from 100 Hz to 500 Hz is much smaller than that of using LC resonance. The efficiency of magnetic induction is lower than that of LC resonance, but it provides a broader bandwidth. {The placement and orientation of the transmit and receive coils can dramatically affect communication performance. In our experiment setup, we let coils face the metal pipe wall. This can provide a strong coupling between the two coils. Depending on the applications, the coils can be placed in other formats  \cite{zangl2010wireless}.}

	\begin{figure}[t]
		\centering
		\includegraphics[width=0.45\textwidth]{fig/resonance}
		\vspace{-5pt}
		\caption{Impedance of circuits with and without LC resonance.}
		\label{fig:resonance}
		\vspace{-10pt}
	\end{figure}
	\section{Communication System Design}
	
	Our design objective is to obtain a through-metal wireless communication system with a reasonable data rate and range to support localization and communication with sensors and robots in metal-constrained environments. In this section, we introduce the system architecture and the detailed design of the transmitter and receiver. 
	\subsection{System Architecture} 
	
	The transmitter uses a microcontroller to generate on-off (return-to-zero) keying modulated signals. The output voltage from the microcontroller is low, which cannot be used to power the transmitter coil. Thus, the signals are amplified by a MOSFET-based switching amplifier. The receiver captures attenuated signals, which cannot be detected by the microcontroller's Analog-to-Digital Converter (ADC). Therefore, an amplifier is used to increase the received signal voltage, and the microcontroller's ADC is used to convert analog signals to digital signals. The sampled digital data is captured and saved in text format by using serial communication monitors (Putty) which is imported into MATLAB for demodulation and BER calculation. 
	
\begin{figure}[t]
		\centering
		\includegraphics[width=0.35\textwidth]{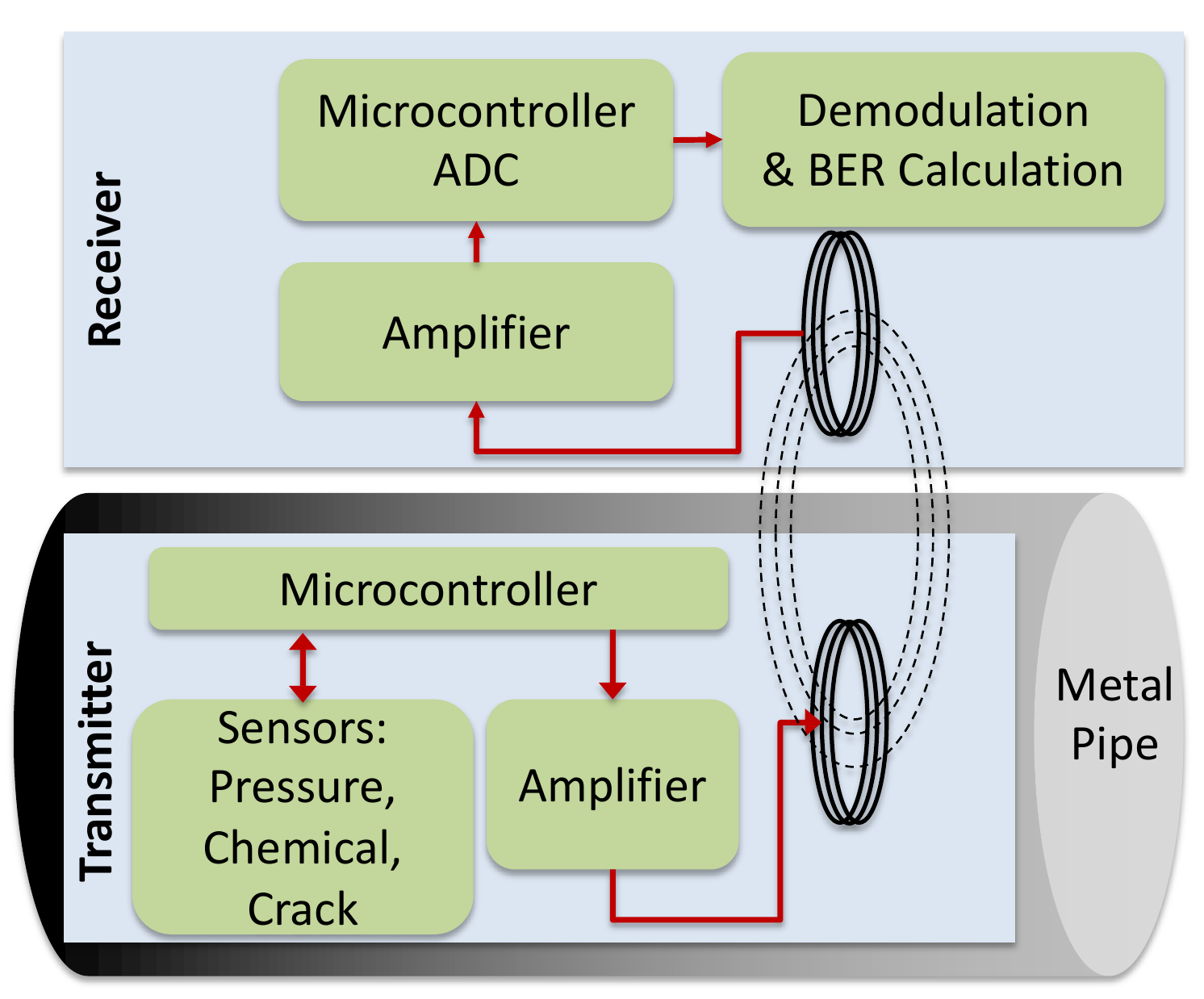}
		\vspace{-5pt}
		\caption{System architecture. The transmitter can also be equipped with a receiving circuit. Similarly, the receiver can transmit with a transmitter circuit, i.e., the communication can be bidirectional.  }
		\label{fig:system}
		\vspace{-0pt}
	\end{figure}

\begin{figure}[t]
		\centering
		\includegraphics[width=0.45\textwidth]{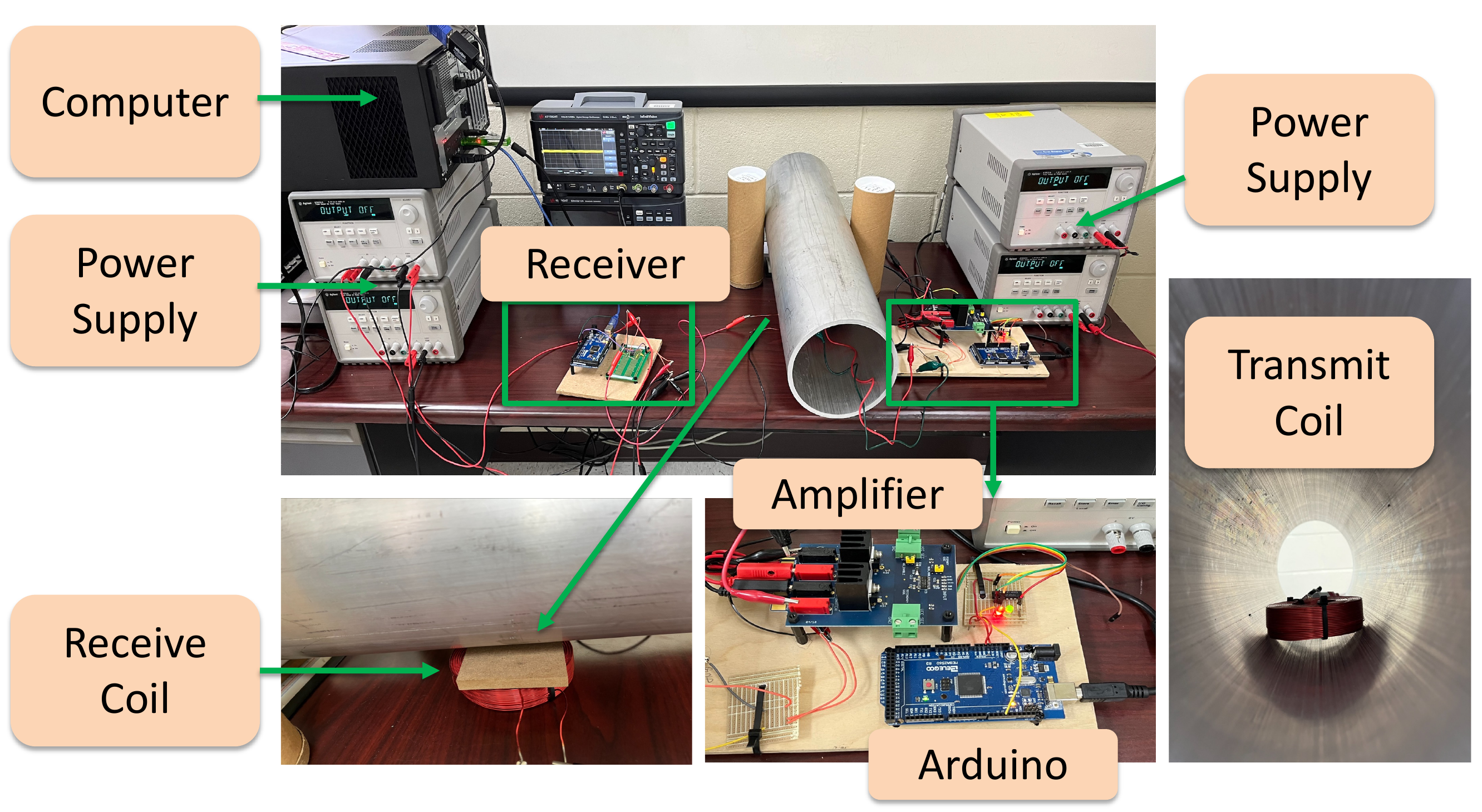}
		\vspace{-5pt}
		\caption{The developed magnetic induction through-metal wireless communication prototype. }
		\label{fig:testbed}
		\vspace{-15pt}
	\end{figure}

	Both the transmitter and the receiver have bottlenecks that limit the data rate and communication range. The following four design challenges need to be addressed.
	\begin{itemize}
		\item Modulation scheme. Carrier modulation is widely used in wireless communication systems. However, in this system, the carrier frequency is lower than several kHz and thus the baseband frequency must be much lower, which significantly reduces the data rates. In this paper, we only use baseband modulation without carrier modulations.
		\item Frequency response of the transmitter amplifier and receiver amplifier. Most Radio Frequency (RF) amplifiers operate at frequency bands higher than 9 kHz. The low signal frequency requires us to identify or design amplifiers for the system. The transmitter amplifier must output high power and the receiver amplifier needs to increase the output voltage. 
		\item Receiver ADC sampling rate.  Since we use low-cost microcontrollers, the ADC has a limited sampling rate. This reduces the signal demodulation accuracy. 
		\item A communication protocol is needed to demodulate received signals. Besides simply transmitting binary 0 and 1, we need a communication protocol to encode data into data packets and evaluate the BER based on received data packets. 
	\end{itemize}
	The system architecture and the designed prototype are shown in Fig.~\ref{fig:system} and Fig.~\ref{fig:testbed}, respectively. Next, we introduce our design of the system, including the transmitter and the receiver. 
	
	\subsection{Transmitter}
	Both the transmitter and the receiver use Arduino Mega 2560 (\$20.99 each) as the microcontroller.  
	\subsubsection{Modulation}
	In this paper, we directly transmit the baseband line coding signals without carrier modulation. In existing wireless systems, baseband signals are modulated using carrier signals mainly because of the following two reasons. First, the baseband frequency is much lower than the carrier frequency, and directly transmitting baseband signals requires large antennas due to the long wavelength. Carrier modulation allows the use of small antennas. Second, carrier modulation can shift the baseband signals to the allowed frequency band to avoid interference and follow spectrum regulations. 
	In this paper, the through-metal communication system does not have these issues. First, using $f_c$ in Equ. \ref{equ:fc} as carrier frequency requires low-frequency baseband signals, i.e., the baseband signal frequency must be much lower than $f_c$. This significantly reduces the achievable data rates. Using $f_c$ as the baseband signal frequency without carrier modulation can achieve a higher data rate. Second, since $f_c$ is only several kHz and most of the off-the-shelf microcontrollers' clocks are faster than 1 MHz, it is not challenging to directly modulate signals at $f_c$. Also, coils can efficiently transmit such low-frequency signals mainly carried by magnetic fields. Since it does not have high efficiency in radiation, the generated electromagnetic fields are only constrained to a small area that cannot propagate. 
	
	This system uses on-off keying with return-to-zero signals. When a binary ``1'' is transmitted, the output signal is a logic high signal followed by a logic low signal. When a binary ``0'' is transmitted, the output signal is constantly low.  The duration of the logic high and logic low signals can be reconfigured. As a result, the transmit symbol rate and frequency are fully programmable. 
	\subsubsection{Amplifier}
	The transmitter amplifier must provide sufficient current to drive the transmit coil. In this system, we use the EiceDRIVER 2EDF7275F Gate Driver development board (\$50.32), which is a voltage-switching amplifier. Two IRF510 MOSFETs are soldered onto the development board. The MOSFET driver requires two PWM input signals (INA and INB) to control the upper and the lower MOSFETs. The microcontroller generates modulated signals and connects to the INA. An inverter is used to invert the modulated signals, which are provided to INB. In our experiment, the amplifier can output power higher than 10 W. We did not test the maximum output power in order to avoid any parts failure.  
	\subsubsection{Communication Data Packets}
	In order to demodulate the transmitted data, we use a sequence with 4 binary ``1'' for synchronization and indication of the start of the packet. Then, each packet is followed by 16 binary numbers. Between each two data packets, we set an interval of 200 $\mu$s without any signal transmission to indicate the end of the previous data packet.  
	\begin{figure}[t]
		\centering
		\includegraphics[width=0.45\textwidth]{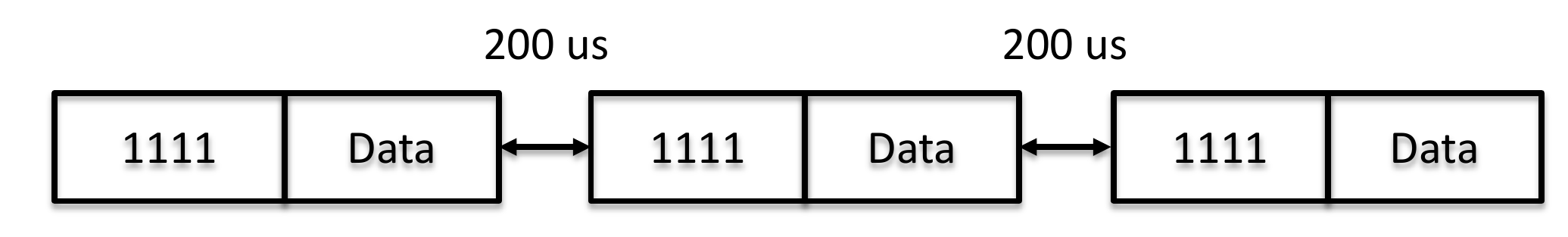}
		\vspace{-5pt}
		\caption{Data packets with 200 $\mu$s intervals. }
		\label{fig:packet}
		\vspace{-10pt}
	\end{figure}
	Note that, synchronization codes, such as Barker code, and error correction codes can be used to improve the effective communication data rates. Since we only use short packets to measure the BER, the error correction codes are not used. Also, the packet start sequence and the packet interval reduce the effective data rate. 
	\subsection{Receiver}
	The receiver unit uses three major components, including the amplifier, ADC in the microcontroller, and demodulation algorithms. 
	
	The received signal is amplified using an LM741 amplifier. Since the signal has a low frequency, such a low-cost amplifier can be used. The received signal is increased to around 1.5~V for sampling. The microcontroller ADC has a reference voltage which is adjusted to ensure that the received signals are not saturated. We set the ADC sampling frequency as 50k samples/second. The sampled data are first saved in the microcontroller, then sent to a computer for offline demodulation through serial ports. 
	
	However, if we send the sampled data via the serial port once the ADC obtained it, this increases the delay, i.e., both the ADC process delay and the serial port communication delay between the microcontroller and the computer affect the sampling rate. Using this approach we can only get around 2000 samples/second. In order to increase the sampling rate, we directly control the hardware and save converted data in a register. The microcontroller has limited memory which cannot save all the samples. As a result, it has to periodically send data through the serial ports and empty its register. This process increases delay and a few samples can be lost. Here, we set the memory length as 2500 samples. Consider the sampling rate, every 50 ms some samples are lost due to serial communication. This will be considered in the demodulation and BER calculation. Note that, if the demodulation is online, i.e., the microcontroller processes the demodulation instead of sending it to the computer via serial ports, this issue can be addressed.

	\subsection{Demodulation and BER Calculation}
	\begin{figure}[t]
		\centering
		\includegraphics[width=0.45\textwidth]{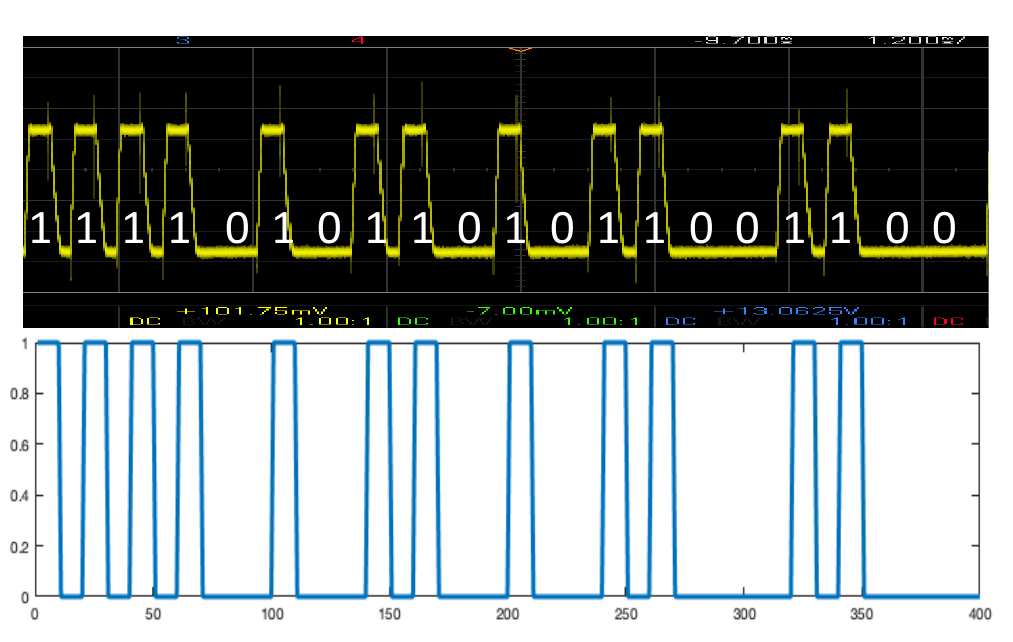}
		\vspace{-5pt}
		\caption{Received waveform and transmitted waveform. The first four bits are the preamble for synchronization and detecting the start of the packet. A binary ``1'' is represented by a high-to-low signal and a binary ``0'' is represented by a low signal.   }
		\label{fig:osc}
		\vspace{-0pt}
	\end{figure}
	The demodulation has the following two challenges. First, the receiver needs to accurately estimate the start of the packet, then use the symbol time to demodulate each symbol.  Second, the missing samples due to the serial communication need to be considered since it introduces extra errors. An example of the received signal is shown in Fig.~\ref{fig:osc}. The symbol rate is 2500 symbols/second. Besides 4 bit ``1111'' preamble, the data has 16 bits (overall, the packet has 20 bits). The upper figure shows the received signal in an oscilloscope. The associated binary number and the transmitted waveform are shown in the lower part of Fig.~\ref{fig:osc}. As we can see, the received waveform is slightly distorted by the coils and amplifiers, i.e., the high-level signal decreases slowly which affects the low-level signal detection. These issues generate impacts on the BER.
	
	Consider the symbol rate is $R_s$ and the sampling rate is $R_m$. Then every symbol has $R_m/R_s$ samples. We preprocess the received data. First, we find the moving average value of 20 symbols. Since the preamble consists of all binary ``1''s, in the packet, the percentage of binary ``1'' is higher than that of binary ``0''. Therefore, the mean value is scaled based on the percentage, which was subtracted from the original data. Then, the mean value of the sampled data is around 0. All the positive values are considered as ``1'' and all the negative values are considered as ``0''. This step can be considered the automatic gain control unit in wireless systems which can output a suitable signal amplitude given various input signal amplitudes.  
	
\begin{figure}[t]
		\centering
		\includegraphics[width=0.45\textwidth]{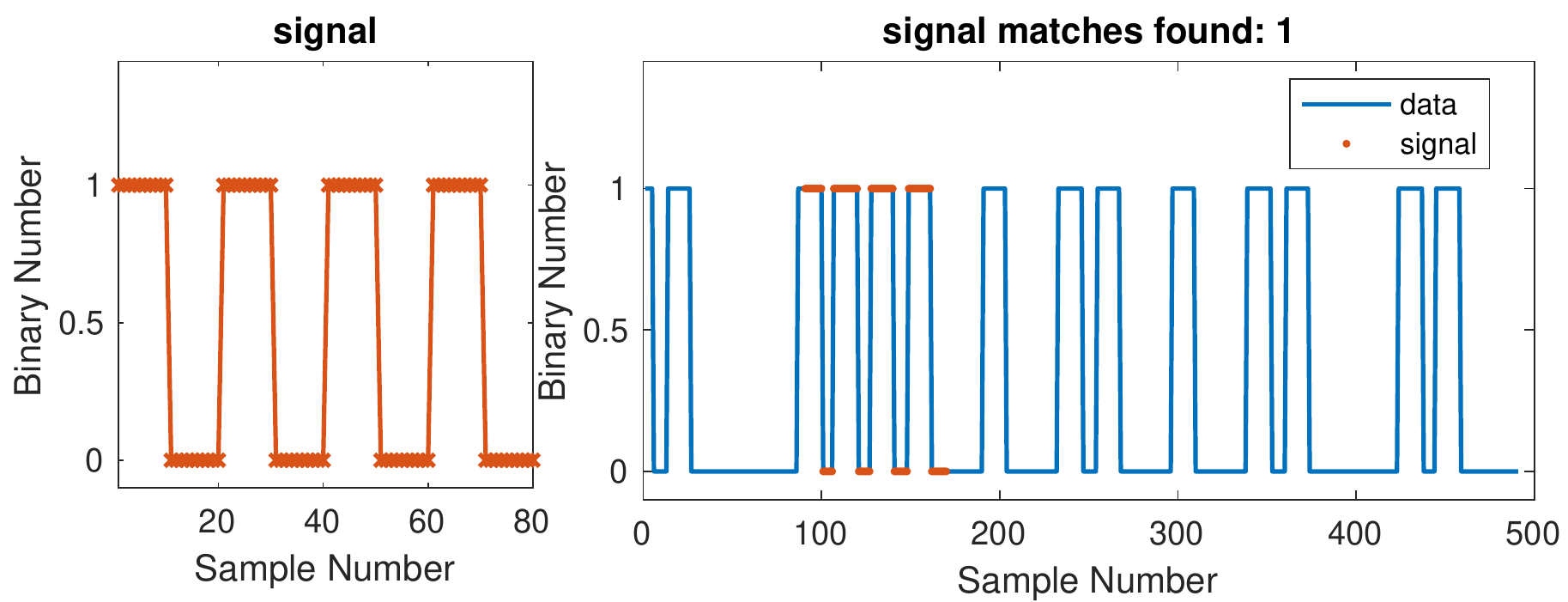}
		\vspace{-5pt}
		\caption{Preamble detection example. (left) preamble signals; (right) detected preamble signals in a sequence. }
		\label{fig:signal_match}
		\vspace{-0pt}
	\end{figure}

	First, we use $4R_m/R_s$ data points to form a data segment of ``1111''. According to the transmission protocol, each data packet has $20R_m/R_s$ samples. We search for the packet preamble in every $24R_m/R_s+10$ samples. This ensures that at least one preamble will be included in the samples. The extra 10 samples are due to the 200 $\mu s$ interval between packets. The squared Euclidean distance between this data segment and the sampled array is obtained by moving from the first sample. The sequence that gives the minimum Euclidean distance is considered the preamble. An example is shown in Fig.~\ref{fig:signal_match}. The left-hand side shows the preamble signals and the right-hand side shows the detected preamble in the received and preprocessed signals. Once the preamble is detected, the rest of the signals are demodulated symbol by symbol. The mean received signal is obtained. If it is higher than a threshold, the symbol is considered binary ``1'', while if it is lower than the threshold, the symbol is considered binary ``0''.
	
	Each data packet carries the same data sequence of ``0101101011001100'', as shown in Fig.~\ref{fig:osc}, to facilitate the calculation of BER. In this way, we have the knowledge of the transmitted packets, upon which we can find the errors. The BER is obtained by using the ratio of the total number of errors over the total number of transmitted bits, which does not include the preamble.  
	
	\section{Performance Evaluation}
	\label{sec:numerical}
	In this section, we first evaluate the communication data rates and the challenges of communication bandwidth without the metallic pipe to understand the system's performance. Then, we measure and discuss the through-metal communication performance by placing the receive coil close to the pipe and the transmit coil in the pipe. 
	\subsection{Symbol Rates and Bandwidth}
	\begin{figure}[t]
		\centering
		\includegraphics[width=0.49\textwidth]{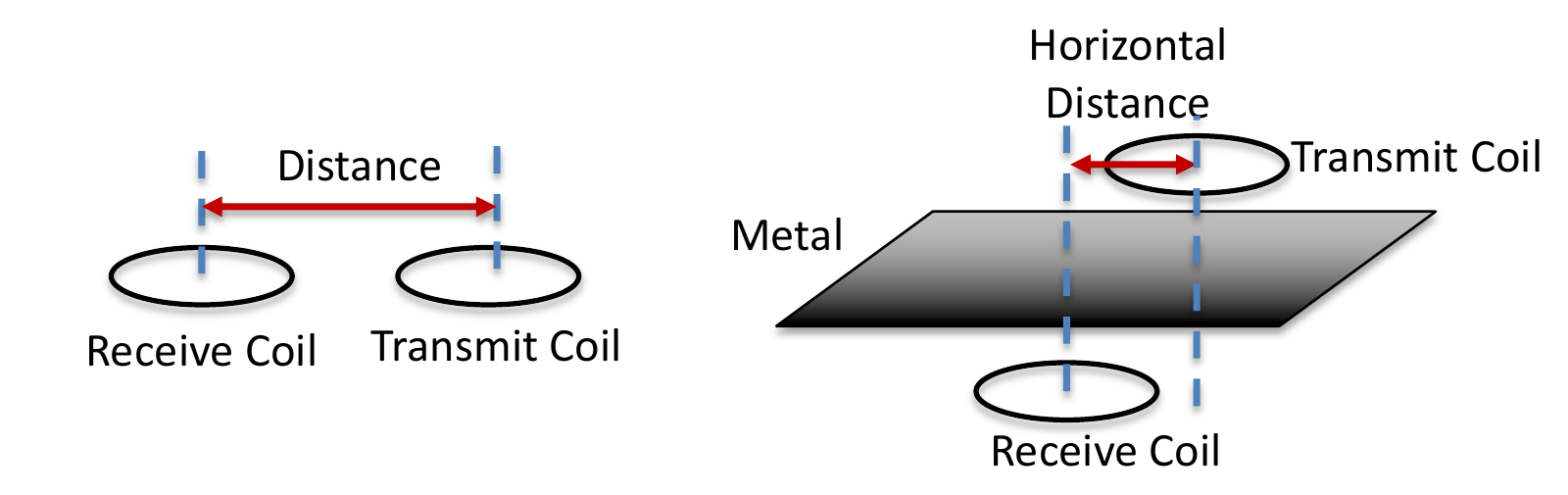}
		\vspace{-5pt}
		\caption{Illustration of the distance between the transmit and receive coils. (Left) coplanar coils in the air without metallic pipes. (Right) the receive coil is outside of the pipe and the transmit coil is in the pipe; the horizontal distance is used.}
		\label{fig:distance}
		\vspace{-0pt}
	\end{figure}

	First, we place the transmit coil and the receive coil coplanar, as shown in Fig.~\ref{fig:distance} (left). Since we use On-Off Keying signals, 1 symbol carries 1 bit. However, the symbol rate is not equivalent to the data rate, because of the preamble and the interval between packets.  
	We consider four symbol rates: 500, 1000, 2500, and 5000 symbols/second.  
	\begin{figure}[t]
		\centering
		\includegraphics[width=0.45\textwidth]{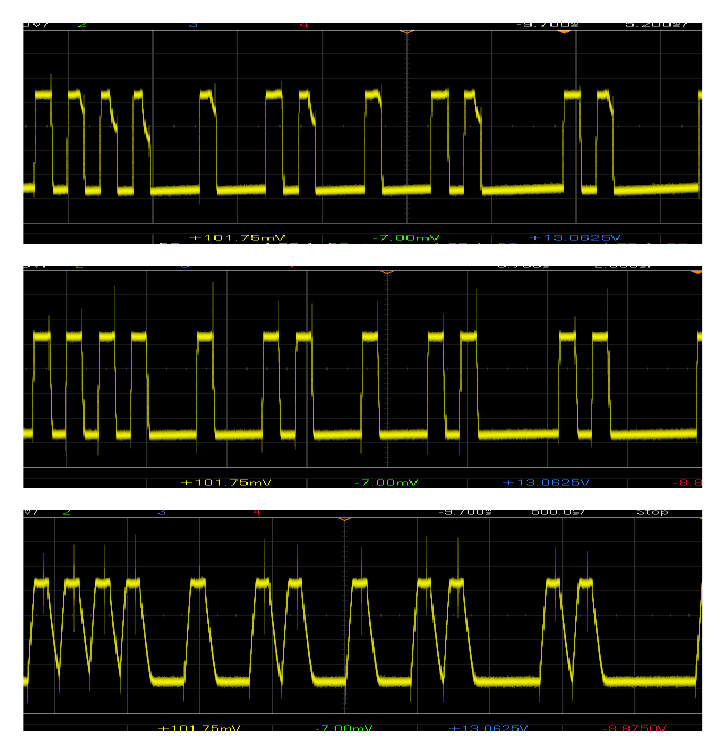}
		\vspace{-5pt}
		\caption{Received signal waveforms for 500 symbols/second (top), 1000 symbols/second (middle), and 5000 symbols/second (bottom). }
		\label{fig:waveform}
		\vspace{-10pt}
	\end{figure}
	
	To understand the impact of data rates and bandwidth, we show the received signal waveforms in Fig.~\ref{fig:waveform} for 500, 1000, and 5000 symbols/second. The received signal waveforms of 2500 symbols/second are shown in Fig.~\ref{fig:osc}. As we can see, for 500 and 1000 samples/second, the received signal waveforms are not distorted and the difference between binary ``1'' and ``0'' is clear. However, for 2500 and 5000 samples/second, the waveforms are getting distorted, especially for 5000 samples/second. The signal of ``1'' decreases slowly from the high level to the low level and a long tail extends to the low level. This can increase the detection error by missing binary ``1'' or considering binary ``0'' as binary ``1''.

	\begin{figure}[t]
		\centering
		\includegraphics[width=0.45\textwidth]{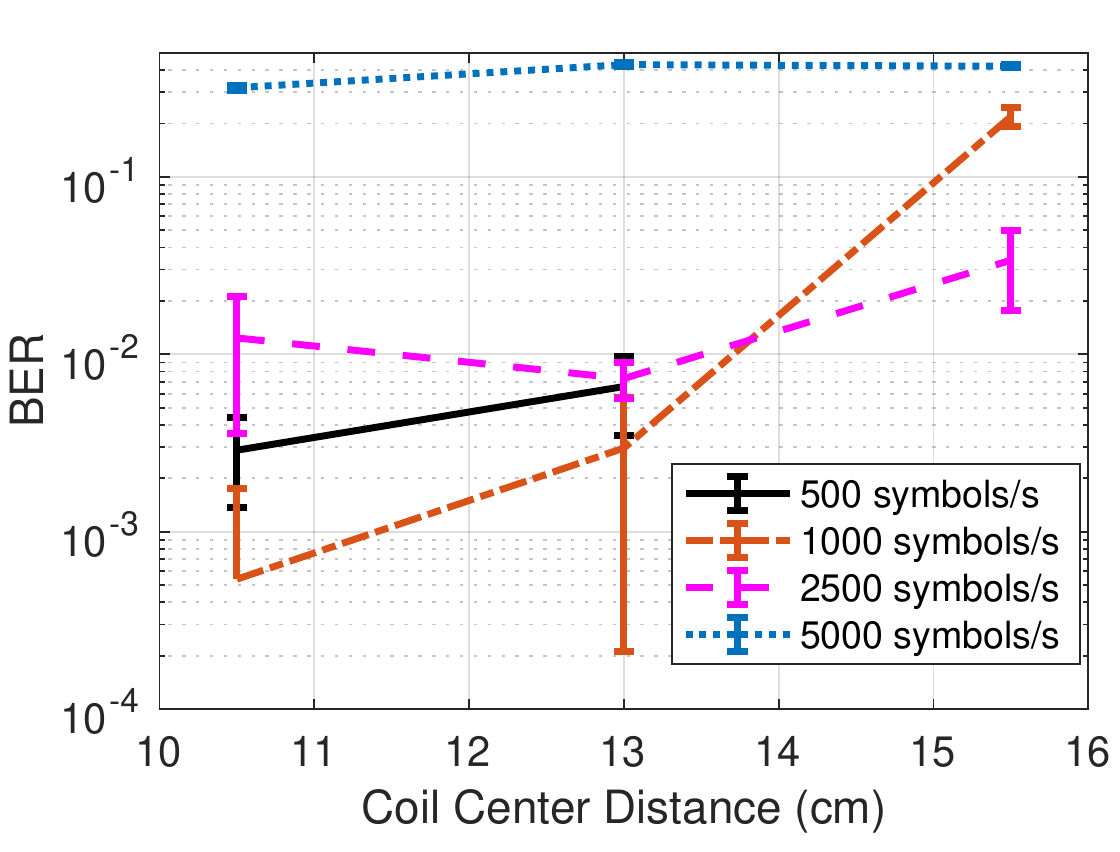}
		\vspace{-5pt}
		\caption{Measured Bit-Error-Rate (BER). }
		\label{fig:BER}
		\vspace{-0pt}
	\end{figure}
	For 500 symbols/second data rate, we reduce the ADC sampling rate from 50k samples/second to 10k samples/second. Otherwise, we receive a large number of redundant samples. The measured BER in the air without the metallic pipe is shown in Fig.~\ref{fig:BER}. We increase the distance between the transmit coil and the receive coil from 10.5 cm to 15.5 cm. As we can see in the figure, 5000 symbols/second has a high BER, which is highly unreliable. This is due to the frequency response of the coil. The impedance of the coil is $j\omega L$, where $j=\sqrt{-1}$ and $L$ is the self-inductance. For 5000 symbols/second, the signal has a broad bandwidth and thus the impedance of the coil changes significantly, which generates drastically different signal strength at different frequencies. Also, the coil has parasitic capacitances that limit its highest operating frequency as an inductor. Note that, the BER of 5000 symbols/second can be improved if more complex demodulation schemes are used. Here, we use the same demodulation scheme for 2500 symbols/second. 
	
	For 500 and 1000 symbols/second, they have low BER when the distance is small. Such low symbol rates allow the signal to be received without distortion. However, when the distance is 15.5 cm, their BER increases significantly. For 500 symbols/second, we cannot even detect the packet and the BER is not shown. This is because the coupling between the transmit and receive coils decreases as the distance increases. For 2500 symbols/second, this effect is weaker because the higher frequency increases the coupling. But, as discussed before, an even higher frequency can distort the signal. As shown in Fig.~\ref{fig:BER}, 2500 symbols/second provide the most reliable performance due to the tradeoffs between the bandwidth and the magnetic coupling.

	\subsection{Through-metal Communication Performance}
	
	\begin{figure}[t]
		\centering
		\includegraphics[width=0.4\textwidth]{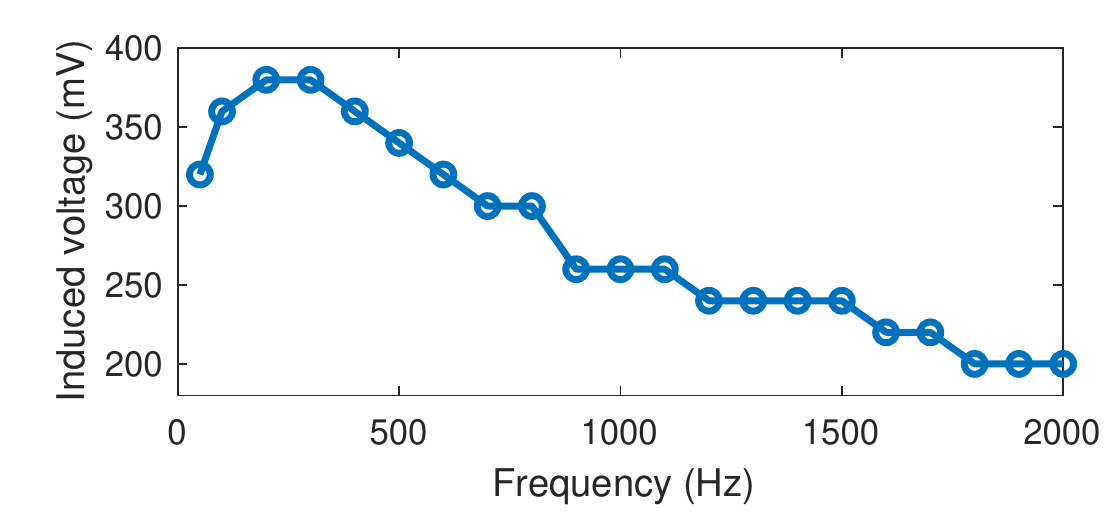}
		\vspace{-5pt}
		\caption{Induced voltage (peak-to-peak) in the receive coil when the peak-to-peak voltage in the transmit coil is 10 V. The input signal is a sine wave. }
		\label{fig:induced_voltage}
		\vspace{-10pt}
	\end{figure}
	
	\begin{figure}[t]
		\centering
		\includegraphics[width=0.45\textwidth]{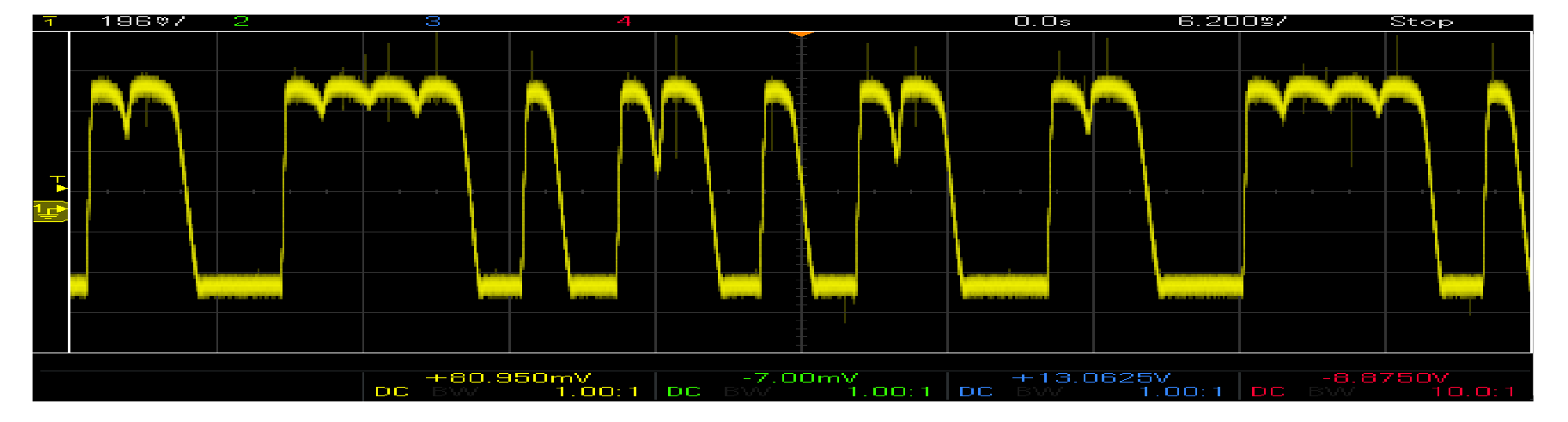}
		\vspace{-5pt}
		\caption{Signal distortion at 500 symbols/second due to the metallic pipe. }
		\label{fig:metal_distortion}
		\vspace{-10pt}
	\end{figure}

	In this subsection, {we place the transmit coil in the metallic pipe and the receive coil outside of the pipe, as shown in Fig.~\ref{fig:distance}. The receive coil's top face is 1 cm under the pipe, and the coil is placed in the middle of the metallic pipe. The position of the transmit coil is shown on the right-hand side of Fig.~\ref{fig:testbed}. }
	
	{ A 10 V peak-to-peak sine signal is given to the transmit coil with frequencies from 50 Hz to 2000 Hz. We measure the induced voltage in the receive coil. As shown in Fig.~\ref{fig:induced_voltage}, the voltage in the receive coil is strong between 200 Hz and 300 Hz, then it gradually drops as the frequency increases. Beyond 500 Hz, the signal drops significantly, and this indicates that using signal bandwidth above 500 Hz can generate strong signal distortions.}
	
	We tested 1000, 2500, and 5000 symbols/second, but the signals are highly distorted and we cannot identify the symbols. An example of the received signal with 500 symbols/second is shown in Fig.~\ref{fig:metal_distortion}. As we can see, the received signals are also distorted, but the on-off still exists. The return-to-zero becomes non-return-to-zero. Therefore, we modify the preamble sequence and demodulate the received signal based on the logic level. The transmission power is around 1.8 W (output voltage 8 V, output current 0.225 A), which is measured by the power supply of the transmit amplifier. 
	
\begin{figure}[t]
		\centering
		\includegraphics[width=0.4\textwidth]{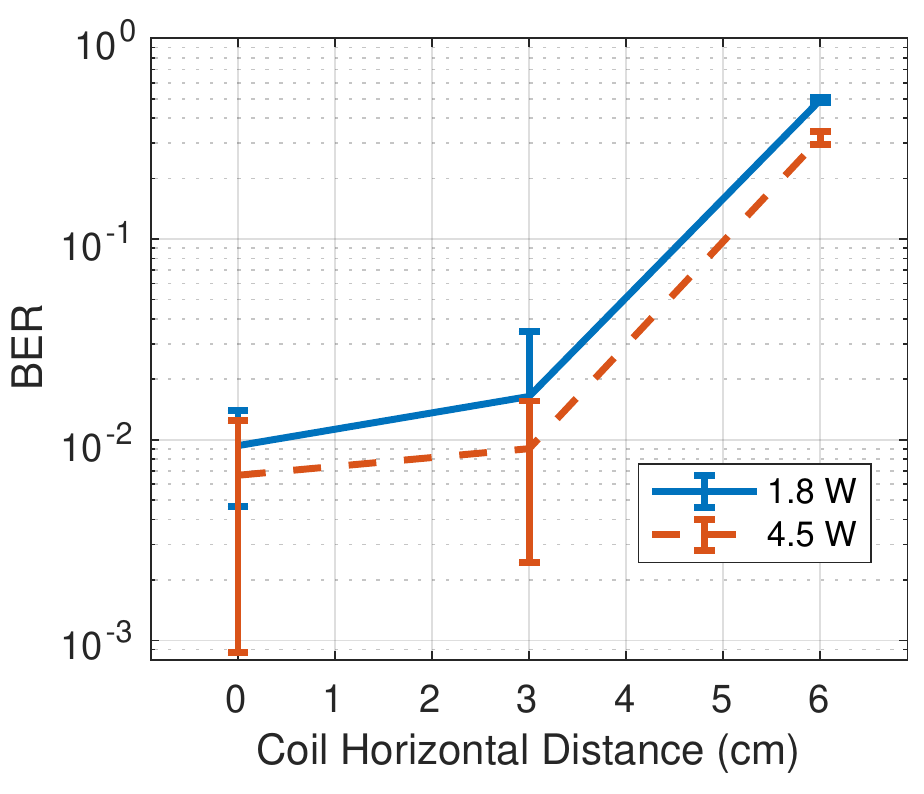}
		\vspace{-5pt}
		\caption{Impact of the transmission power. }
		\label{fig:metal_ber}
		\vspace{-0pt}
	\end{figure}

	Note that, the transmission power varies around 1.8 W, which is not a constant due to the different amplitudes of the binary ``1'' and ``0''. We also increase the transmission power to around 4.5 W (output voltage 15 V, output current 0.3 A). The measured BER is shown in Fig.~\ref{fig:metal_ber}. { It should be noted that such a high power consumption prevents us from using typical wireless communication protocols. The through-metal communication system in this paper can be used for low-volume data transmission applications, such as sensing, localization, and in-pipe robotic sensor tracking. Novel communication protocols that can minimize energy consumption are required.}
	
	\begin{figure}[t]
		\centering
		\includegraphics[width=0.5\textwidth]{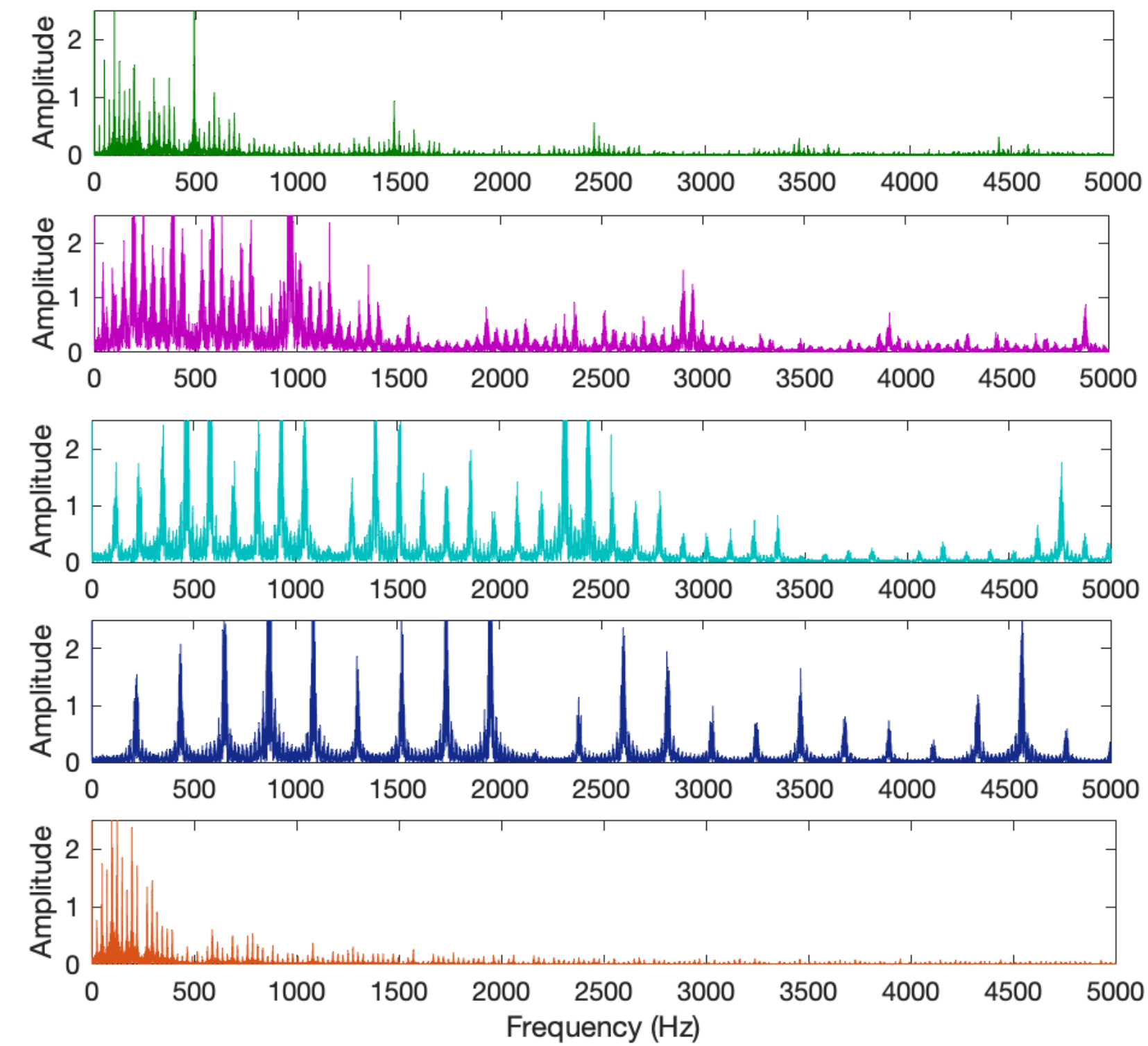}
		\vspace{-5pt}
		\caption{Spectrum of the received signals. From the top to bottom: (1) 500 symbols/second without the metallic pipe, (2) 1000 symbols/second without the metallic pipe, (3) 2500 symbols/second without the metallic pipe, (4) 5000 symbols/second without the metallic pipe, and (5) 500 symbols/second for through-metal communication. }
		\label{fig:spectrum}
		\vspace{-10pt}
	\end{figure}
	The distance between the transmit coil and the receive coil is determined by both the vertical distance between the pipe and the receive coil and the horizontal distance, as shown in Fig.~\ref{fig:distance}, which cannot directly reflect the transmit coil's location in the pipe. {Here, we only consider the horizontal distance, which is increased from 0 cm to 6 cm.} As we can see in Fig.~\ref{fig:metal_ber}, as the horizontal distance increases, the high transmission power can always effectively reduce the mean BER by 36\% on average. 
	\subsection{Optimal Frequency}
	In Section~\ref{sec:optimal}, we find the optimal frequency should be lower than 3.7 kHz. However, when we place the transmit coil in the pipe, we notice that the maximum symbol rate is around 500 symbols/second. If we only transmit binary ``1'', such a symbol rate generates 500 Hz square waves. The difference between the achievable symbol rate and the optimal frequency is due to the change in coil impedance. As shown in Fig.~\ref{fig:coilparameter}, the coil self-inductance reduces, and its resistance increases. This impact increases as the frequency becomes higher. For example, in Fig.~\ref{fig:coilparameter}, the coil self-inductance and resistance change more significantly at 1 kHz than that at 100 Hz. As a result, besides the skin depth impact, the coil impedance change further reduces the optimal frequency. 
	
	Also, we compare the received signal spectrum in Fig.~\ref{fig:spectrum}. As we can see, using 500 symbols/second, most of the power is within 500 Hz, while higher symbol rates increase the bandwidth significantly, which experiences more distortion when the coil is placed in or close to the metallic pipe. The prototype transmits at 500 symbols/s and the achieved BER is slightly lower than 0.01 when the distance is smaller than 3 cm and the transmission power is 4.5 W, as shown in Fig.~\ref{fig:metal_ber}. Under such conditions, the Euclidean distance between the transmit coil and the receive coil is around 3.4 cm, which includes both the vertical distance and the horizontal distance. Theoretically, the achievable data rate is 500 bps. However, due to the errors, preamble, and interval between packets, the effective data rate is slightly lower. Optimizing the communication protocols and using error detection and correction codes can effectively address these issues. 
	\section{Conclusion}
	Sensors in metal-constrained environments, such as toxic liquid/gas containers, oil and gas transportation pipelines, and space vehicles, play an important role in gathering critical information. However, metal walls can block Radio Frequency (RF) signals due to their high conductivity. Wireless communication for sensors in such environments is challenging. This paper develops a low-cost through-metal wireless communication system that can penetrate Aluminum metallic pipe with 7 mm thickness. The transmitter and the receiver use coils for magnetic induction communication. The prototype is fully reconfigurable, and the modulation waveform and data rates are programmable. We evaluate different symbol rates and transmission power. The On-Off Keying modulation is used, and an average Bit-Error-Rate (BER) of 0.01 with 500 symbols/second can be obtained. Although the developed system is used for wireless communication, it can also be used for wireless sensings, such as metal thickness sensing based on the received signal strength and frequency response. 
	\bibliographystyle{IEEEtran}
	\bibliography{ref}

\begin{thebibliography}{10}
\providecommand{\url}[1]{#1}
\csname url@samestyle\endcsname
\providecommand{\newblock}{\relax}
\providecommand{\bibinfo}[2]{#2}
\providecommand{\BIBentrySTDinterwordspacing}{\spaceskip=0pt\relax}
\providecommand{\BIBentryALTinterwordstretchfactor}{4}
\providecommand{\BIBentryALTinterwordspacing}{\spaceskip=\fontdimen2\font plus
\BIBentryALTinterwordstretchfactor\fontdimen3\font minus
  \fontdimen4\font\relax}
\providecommand{\BIBforeignlanguage}[2]{{%
\expandafter\ifx\csname l@#1\endcsname\relax
\typeout{** WARNING: IEEEtran.bst: No hyphenation pattern has been}%
\typeout{** loaded for the language `#1'. Using the pattern for}%
\typeout{** the default language instead.}%
\else
\language=\csname l@#1\endcsname
\fi
#2}}
\providecommand{\BIBdecl}{\relax}
\BIBdecl

\bibitem{sun2011mise}
Z.~Sun, P.~Wang, M.~C. Vuran, M.~A. Al-Rodhaan, A.~M. Al-Dhelaan, and I.~F.
  Akyildiz, ``Mise-pipe: Magnetic induction-based wireless sensor networks for
  underground pipeline monitoring,'' \emph{Ad Hoc Networks}, vol.~9, no.~3, pp.
  218--227, 2011.

\bibitem{qi2010wireless}
H.~Qi, J.~Ye, X.~Zhang, and H.~Chen, ``Wireless tracking and locating system
  for in-pipe robot,'' \emph{Sensors and Actuators A: Physical}, vol. 159,
  no.~1, pp. 117--125, 2010.

\bibitem{wanasinghe2020digital}
T.~R. Wanasinghe, L.~Wroblewski, B.~K. Petersen, R.~G. Gosine, L.~A. James,
  O.~De~Silva, G.~K. Mann, and P.~J. Warrian, ``Digital twin for the oil and
  gas industry: Overview, research trends, opportunities, and challenges,''
  \emph{IEEE access}, vol.~8, pp. 104\,175--104\,197, 2020.

\bibitem{pelkner2018benefits}
M.~Pelkner, R.~Stegemann, N.~Sonntag, R.~Pohl, and M.~Kreutzbruck, ``Benefits
  of gmr sensors for high spatial resolution ndt applications,'' in \emph{AIP
  Conference Proceedings}, vol. 1949, no.~1.\hskip 1em plus 0.5em minus
  0.4em\relax AIP Publishing LLC, 2018, p. 040001.

\bibitem{el2011vison}
S.~El~Kahi, D.~Asmar, A.~Fakih, J.~Nieto, and E.~Nebot, ``A vison-based system
  for mapping the inside of a pipe,'' in \emph{2011 IEEE International
  Conference on Robotics and Biomimetics}.\hskip 1em plus 0.5em minus
  0.4em\relax IEEE, 2011, pp. 2605--2611.

\bibitem{alnaimi2015design}
F.~B.~I. Alnaimi, A.~A. Mazraeh, K.~Sahari, K.~Weria, and Y.~Moslem, ``Design
  of a multi-diameter in-line cleaning and fault detection pipe pigging
  device,'' in \emph{2015 IEEE International Symposium on Robotics and
  Intelligent Sensors (IRIS)}.\hskip 1em plus 0.5em minus 0.4em\relax IEEE,
  2015, pp. 258--265.

\bibitem{jiao2016monitoring}
S.~Jiao, L.~Cheng, X.~Li, P.~Li, and H.~Ding, ``Monitoring fatigue cracks of a
  metal structure using an eddy current sensor,'' \emph{EURASIP Journal on
  Wireless Communications and Networking}, vol. 2016, no.~1, pp. 1--14, 2016.

\bibitem{pozo2016continuous}
A.~M. Pozo, F.~P{\'e}rez-Oc{\'o}n, and O.~Rabaza, ``A continuous liquid-level
  sensor for fuel tanks based on surface plasmon resonance,'' \emph{Sensors},
  vol.~16, no.~5, p. 724, 2016.

\bibitem{fan2015large}
G.~Fan, Y.~Shen, X.~Hao, Z.~Yuan, and Z.~Zhou, ``Large-scale wireless
  temperature monitoring system for liquefied petroleum gas storage tanks,''
  \emph{Sensors}, vol.~15, no.~9, pp. 23\,745--23\,762, 2015.

\bibitem{kar2021passive}
B.~Kar, T.~Schaechtle, S.~J. Rupitsch, and U.~Wallrabe, ``Passive ultrasonic
  temperature measurement through a metal wall,'' in \emph{2021 IEEE
  Sensors}.\hskip 1em plus 0.5em minus 0.4em\relax IEEE, pp. 1--4.

\bibitem{shan2020developing}
X.~Shan, X.~Li, D.~R. Jackson, and J.~Chen, ``Developing through metal wireless
  sensors in extreme conditions for petroleum industrial applications,'' in
  \emph{2020 IEEE Texas Symposium on Wireless and Microwave Circuits and
  Systems (WMCS)}.\hskip 1em plus 0.5em minus 0.4em\relax IEEE, 2020, pp. 1--6.

\bibitem{guo2019reliable}
H.~Guo and K.~D. Song, ``Reliable through-metal wireless communication using
  magnetic induction,'' \emph{IEEE Access}, vol.~7, pp. 115\,428--115\,439,
  2019.

\bibitem{yan2020deep}
Y.~Yan, D.~Liu, B.~Gao, G.~Tian, and Z.~Cai, ``A deep learning-based ultrasonic
  pattern recognition method for inspecting girth weld cracking of gas
  pipeline,'' \emph{IEEE Sensors Journal}, vol.~20, no.~14, pp. 7997--8006,
  2020.

\bibitem{yang2021bilstm}
L.~Yang and Q.~Zhao, ``A bilstm based pipeline leak detection and disturbance
  assisted localization method,'' \emph{IEEE Sensors Journal}, vol.~22, no.~1,
  pp. 611--620, 2021.

\bibitem{waleed2018pipe}
D.~Waleed, S.~H. Mustafa, S.~Mukhopadhyay, M.~F. Abdel-Hafez, M.~A.~K. Jaradat,
  K.~R. Dias, F.~Arif, and J.~I. Ahmed, ``An in-pipe leak detection robot with
  a neural-network-based leak verification system,'' \emph{IEEE Sensors
  Journal}, vol.~19, no.~3, pp. 1153--1165, 2018.

\bibitem{ashdown2018high}
J.~D. Ashdown, L.~Liu, G.~J. Saulnier, and K.~R. Wilt, ``High-rate ultrasonic
  through-wall communications using mimo-ofdm,'' \emph{IEEE Transactions on
  Communications}, vol.~66, no.~8, pp. 3381--3393, 2018.

\bibitem{erel2021comprehensive}
M.~Z. Erel, K.~C. Bayindir, M.~T. Aydemir, S.~K. Chaudhary, and J.~M. Guerrero,
  ``A comprehensive review on wireless capacitive power transfer technology:
  Fundamentals and applications,'' \emph{IEEE Access}, 2021.

\bibitem{yang2015through}
D.-X. Yang, Z.~Hu, H.~Zhao, H.-F. Hu, Y.-Z. Sun, and B.-J. Hou,
  ``Through-metal-wall power delivery and data transmission for enclosed
  sensors: A review,'' \emph{Sensors}, vol.~15, no.~12, pp. 31\,581--31\,605,
  2015.

\bibitem{zangl2010wireless}
H.~Zangl, A.~Fuchs, T.~Bretterklieber, M.~J. Moser, and G.~Holler, ``Wireless
  communication and power supply strategy for sensor applications within closed
  metal walls,'' \emph{IEEE Transactions on Instrumentation and Measurement},
  vol.~59, no.~6, pp. 1686--1692, 2010.

\bibitem{romero2021miniature}
J.~M. Romero-Arguello, A.-V. Pham, C.~S. Gardner, and B.~Funsten, ``Miniature
  coil design for through metal wireless power transfer,'' in \emph{2021 IEEE
  Wireless Power Transfer Conference (WPTC)}.\hskip 1em plus 0.5em minus
  0.4em\relax IEEE, 2021, pp. 1--4.

\bibitem{yamakawa2014wireless}
M.~Yamakawa, Y.~Mizuno, J.~Ishida, K.~Komurasaki, and H.~Koizumi, ``Wireless
  power transmission into a space enclosed by metal walls using magnetic
  resonance coupling,'' \emph{Wireless Engineering and Technology}, vol. 2014,
  2014.

\bibitem{van2017development}
C.~Van~Pham, A.-V. Pham, and C.~S. Gardner, ``Development of helical circular
  coils for wireless through-metal inductive power transfer,'' in \emph{2017
  IEEE Wireless Power Transfer Conference (WPTC)}.\hskip 1em plus 0.5em minus
  0.4em\relax IEEE, 2017, pp. 1--3.

\bibitem{website}
\BIBentryALTinterwordspacing
 [Online]. Available:
  \url{https://github.com/CCCS-Team/through-metal-communication}
\BIBentrySTDinterwordspacing

\bibitem{liu2021data}
G.~Liu, ``Data collection in mi-assisted wireless powered underground sensor
  networks: directions, recent advances, and challenges,'' \emph{IEEE
  Communications Magazine}, vol.~59, no.~4, pp. 132--138, 2021.

\end{thebibliography}
	
\end{document}